\documentclass[runningheads]{llncs}
\usepackage[T1]{fontenc}
\usepackage{graphicx}
\usepackage{url}
\usepackage{multirow}
\usepackage{float}
\begin{document}
\title{Rethinking Histology Slide Digitization Workflows for Low-Resource Settings}
\titlerunning{Rethinking Histology Slide Digitization Workflows for Limited-Resource Settings}
%
\author{Talat Zehra\inst{1}* \and Joseph Marino\inst{2}* \and Wendy Wang\inst{2}* \and Grigoriy Frantsuzov\inst{2}* \and Saad Nadeem\inst{2}}

\authorrunning{Zehra et al.}
%

\institute{Jinnah Sindh Medical University, Karachi, Pakistan \and
Memorial Sloan Kettering Cancer Center, New York, USA \\
 {*}Equal Contribution. Email: nadeems@mskcc.org}
\maketitle              
\begin{abstract}
Histology slide digitization is becoming essential for telepathology (remote consultation), knowledge sharing (education), and using the state-of-the-art artificial intelligence algorithms (augmented/automated end-to-end clinical workflows). However, the cumulative costs of digital multi-slide high-speed brightfield scanners, cloud/on-premises storage, and personnel (IT and technicians) make the current slide digitization workflows out-of-reach for limited-resource settings, further widening the health equity gap; even single-slide manual scanning commercial solutions are costly due to hardware requirements (high-resolution cameras, high-spec PC/workstation, and support for only high-end microscopes). In this work, we present a new cloud slide digitization workflow for creating scanner-quality whole-slide images (WSIs) from uploaded low-quality videos, acquired from cheap and inexpensive microscopes with built-in cameras. Specifically, we present a pipeline to create stitched WSIs while automatically deblurring out-of-focus regions, upsampling input 10X images to 40X resolution, and reducing brightness/contrast and light-source illumination variations. We demonstrate the WSI creation efficacy from our workflow on World Health Organization-declared neglected tropical disease, Cutaneous Leishmaniasis (prevalent only in the poorest regions of the world and only diagnosed by sub-specialist dermatopathologists, rare in poor countries), as well as other common pathologies on core biopsies of breast, liver, duodenum, stomach and lymph node. The code and pretrained models will be accessible via our GitHub (\url{https://github.com/nadeemlab/DeepLIIF}), and the cloud platform will be available at \url{https://deepliif.org} for uploading microscope videos and downloading/viewing WSIs with shareable links (no sign-in required) for telepathology and knowledge sharing. 

\end{abstract}
\section{Introduction}

Digitizing histology slides is becoming essential for remote consultation (e.g. rural/remote general pathologists seeking opinion from sub-specialist pathologists located in distant urban centers), knowledge sharing (e.g. fast sharing of rare/regional disease histology presentation with national and international colleagues for education), and assistive/augmented artificial intelligence analysis (e.g. computational algorithm development/testing as well as for external validation of developed algorithms on cohorts from different demographics/settings). These digitization workflows however are extremely costly to setup/operate (with digital multi-slide scanners, scanning space, digital cloud/on-premises storage, and IT personnel/technicians) and hence out-of-reach for limited-resource hospitals in low- and middle-income countries where two-thirds of the population reside with a declining number of pathologists (pathologist:population ratio as low as 1:10million). 

To alleviate some of the costs associated with automatic digital multi-slide scanners (costs > \$70,000 \cite{farahani2015whole} and contains precise automatic motor stage and illumination control for capturing/stitching image tiles at 20X or 40X magnification to create whole-slide images [WSIs]), manual single-slide stitching software (> \$3,500) have been released commercially for offline stitching from manual slide scans at 10X via a microscope camera's LIVE video stream. However, these single-slide commercial offline solutions (such as Microvisioneer's mvSlide, Meyer Instrument's Panoptiq, and Promicra's PRO-SCAN) only support specific high-end microscopes (> \$5,000) with high-resolution digital cameras (> \$500) and require expensive high-spec workstations (> \$1,000) to operate. The 10X manual scan constraint is due to requiring a large enough spatial context to avoid local/global stitching drift (common at 20X or 40X \cite{krishna2021gloflow,pellikka2021robust} because of repeating patterns and sparse features) and relatively fast scanning time (1-2 mins for 10X vs 20mins or more for 20X or 40X). 

In this paper, we present a first free open-access cloud-based slide digitization workflow to create, view, and share scanner-quality stitched whole-slide images from uploaded 10X manual scan videos, acquired from cheap (\$200) microscopes with built-in digital cameras. We also present a new deep learning algorithm for deblurring out-of-focus regions, upsampling stitched images from 10X to 40X, and reducing stain brightness/contrast and light-source illumination variations. Finally, we demonstrate our WSI creation efficacy on a World Health Organization-declared neglected tropical disease, Cutaneous Leishmaniasis (prevalent only in the poorest regions of the world and diagnosed by sub-specialist dermatopathologists, residing in urban centers), as well as on core biopsies from other common pathologies of breast, liver, duodenum, stomach and lymph node.

\section{Related Work}
There have been prior works \cite{krishna2021gloflow,pellikka2021robust} exploring whole-slide image creation from videos or image tiles acquired at 20X or 40X but these are focused on automatic whole-slide image creation via a motorized XY stage (e.g. Grundium Ocus20 or Ocus40 automatic single-slide scanners that cost ~\$14,000) and used video data simulated from scanned images (ignoring all the light-source illumination variations common in real manual scan videos). Manual scanning at 20X or 40X is not feasible due to the time it takes to complete the scan (15 mins or more depending on the sample size) and the pairwise image registration errors it can accumulate due to the repeated/sparse features with minimal spatial context. In contrast, we create WSIs from 10X with large spatial context and low scanning time while handling blurring and staining/light-source illumination variations as well as recovering 40X details from 10X input. 

For deblurring, upsampling, and reducing stain variations, Restore-GAN was recently introduced \cite{rong2023enhanced} which used the standard pix2pix architecture with input/output data that again is simulated (added simple Gaussian blur, downsampling, artificial brightness/contrast variations) from scanned whole-slide images (ignoring light-source illumination and more complex non-Gaussian blurring common in real data). In our work, we use real microscope-derived video data co-registered with scanned whole-slide images to simultaneously deblur real out-of-focus regions, upsample 10X to 40X resolution, and reduce staining brightness/contrast as well as light-source illumination variations in stitched WSIs from manual scan video input.

\begin{figure*}[th!]
\begin{center}
\setlength{\tabcolsep}{1pt}
\begin{tabular}{cccc}
Frames & Stitched WSI & Translated WSI & Scanned WSI\\
\multirow{2}{*}{\includegraphics[width=0.20\textwidth]{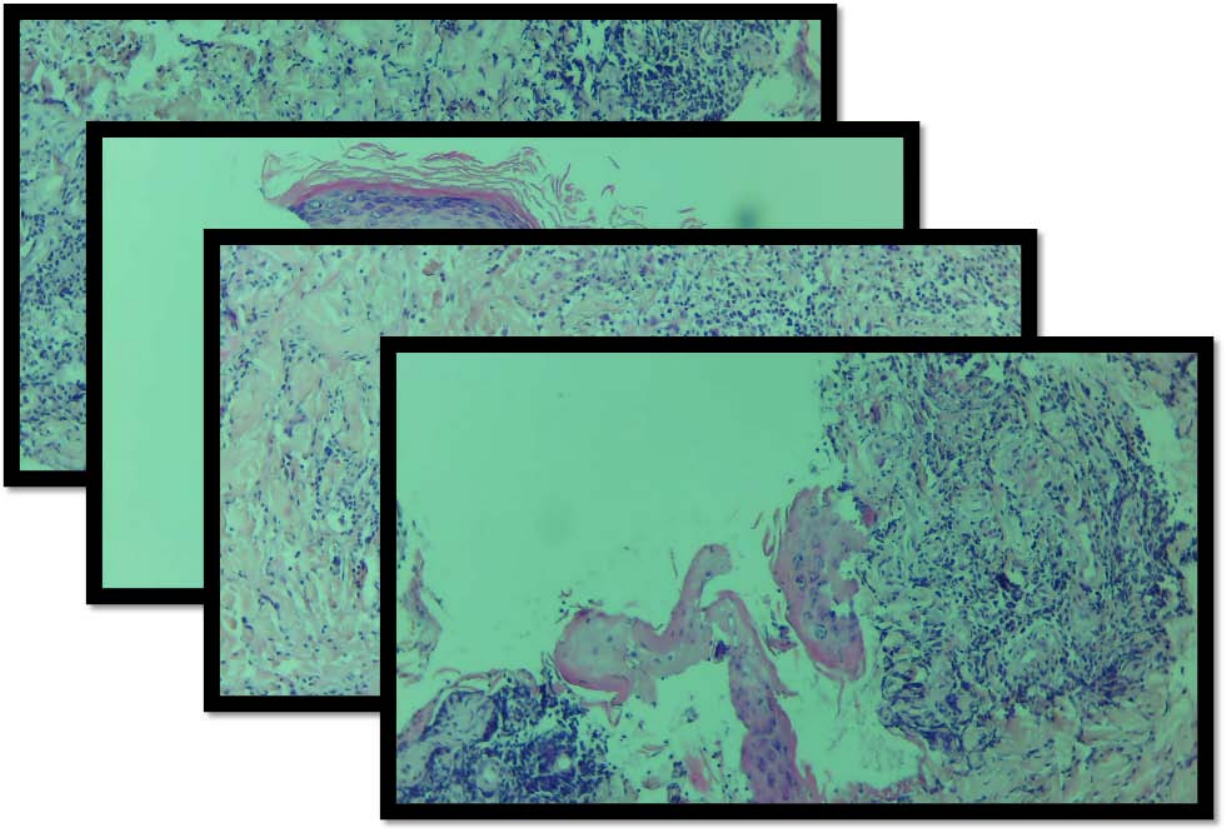}}&
\includegraphics[width=0.26\textwidth]{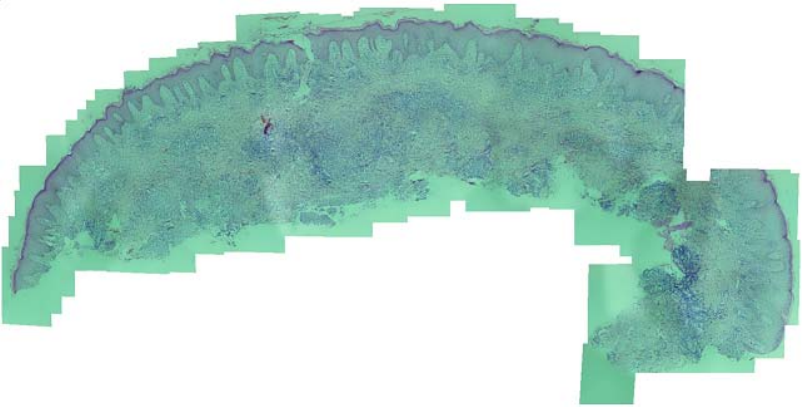}&
\includegraphics[width=0.26\textwidth]{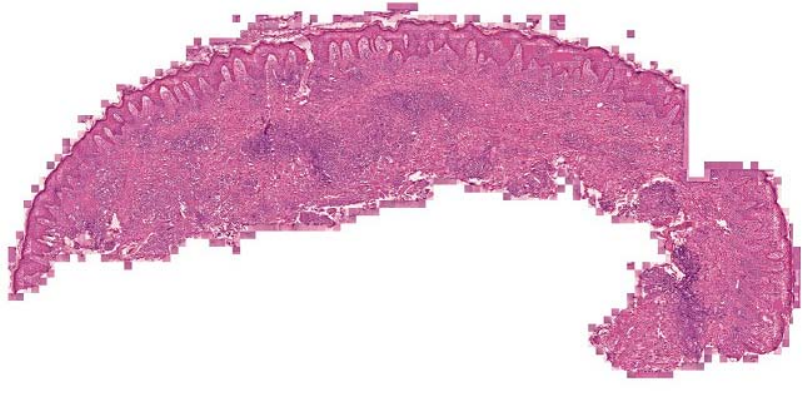}&
\includegraphics[width=0.26\textwidth]{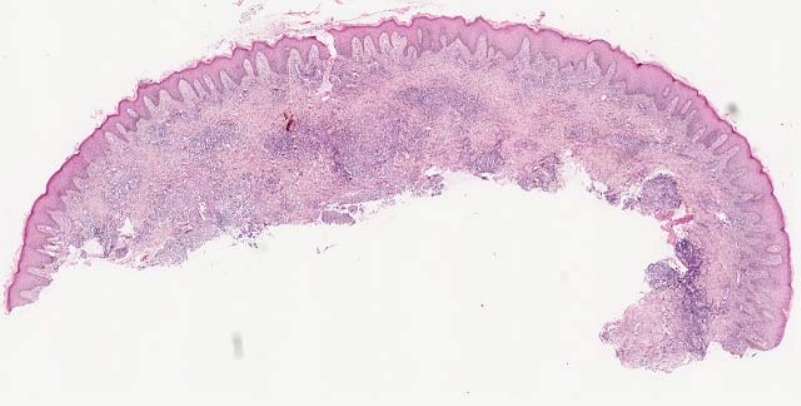}\\
&
\includegraphics[width=0.26\textwidth]{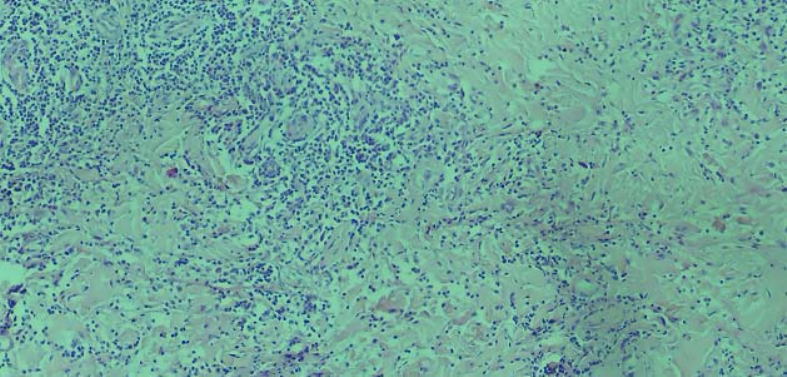}&
\includegraphics[width=0.26\textwidth]{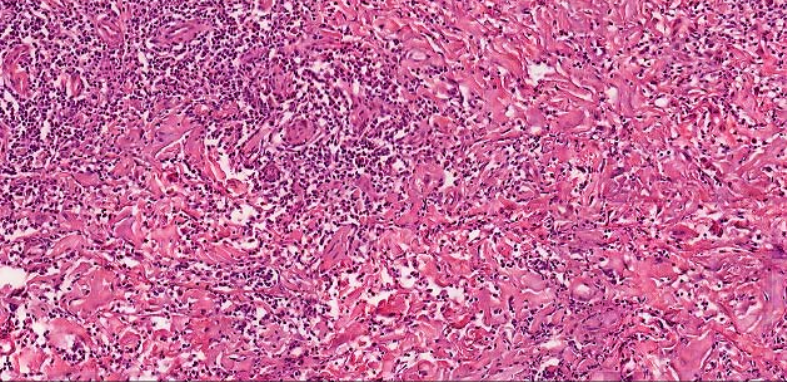}&
\includegraphics[width=0.26\textwidth]{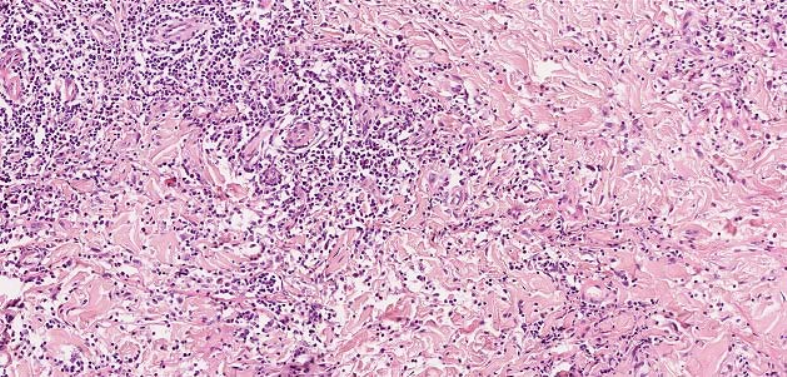}\\
\multirow{2}{*}{\includegraphics[width=0.20\textwidth]{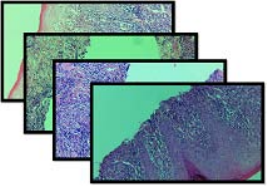}}&
\includegraphics[width=0.2\textwidth]{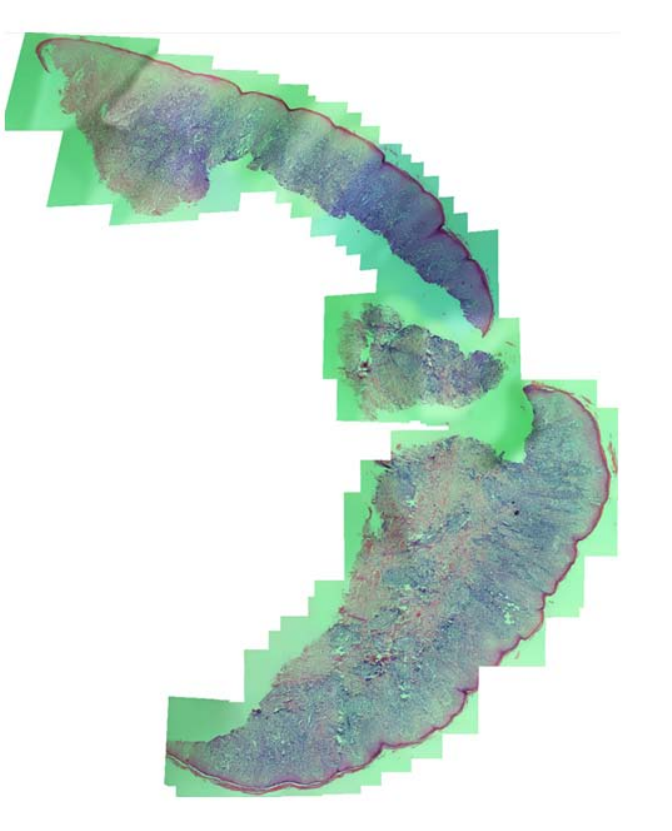}&
\includegraphics[width=0.2\textwidth]{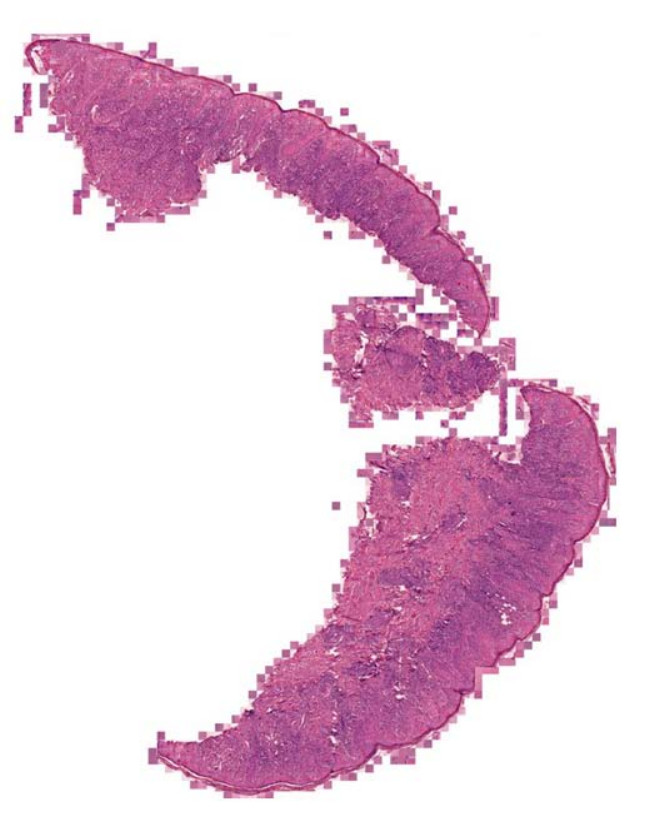}&
\includegraphics[width=0.2\textwidth]{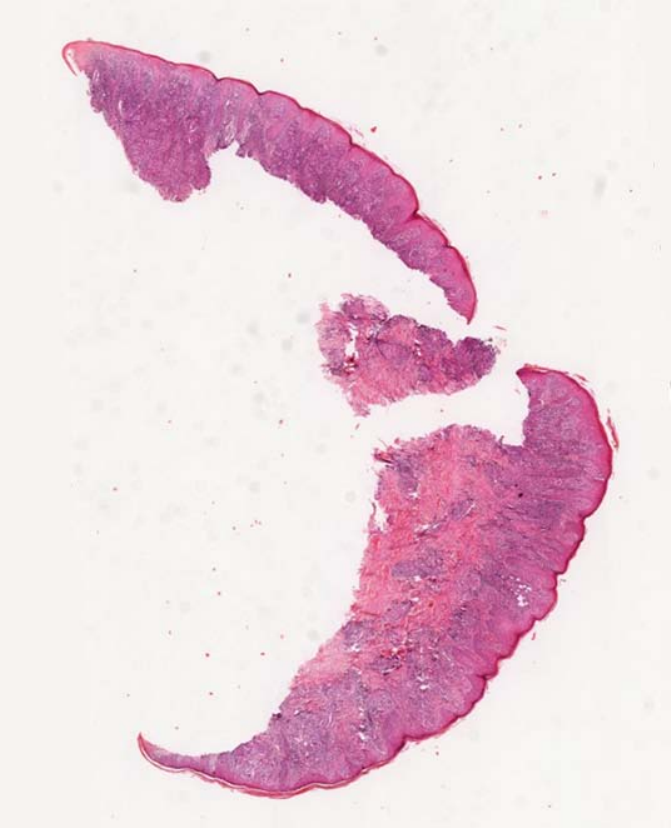}\\
&
\includegraphics[width=0.26\textwidth]{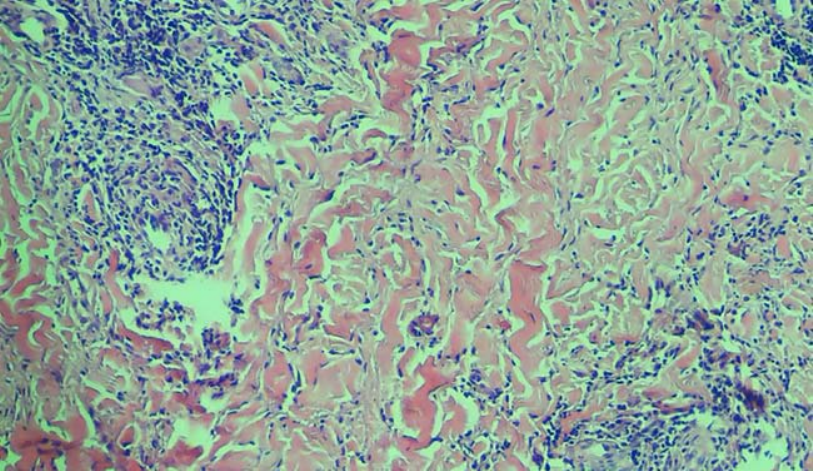}&
\includegraphics[width=0.26\textwidth]{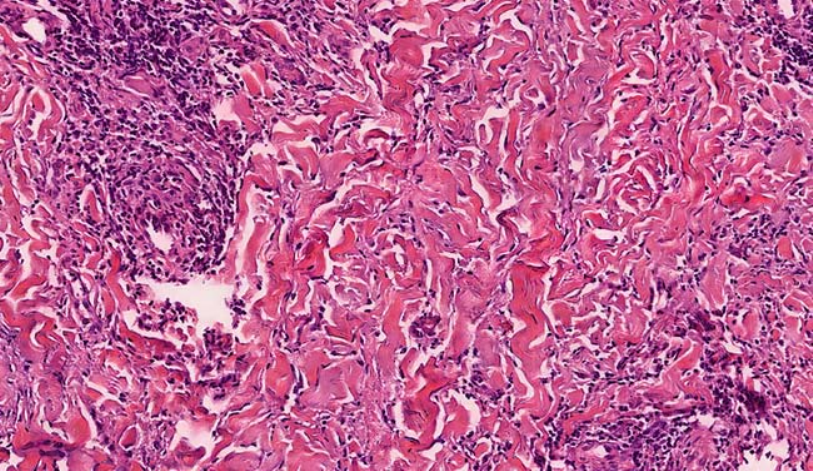}&
\includegraphics[width=0.26\textwidth]{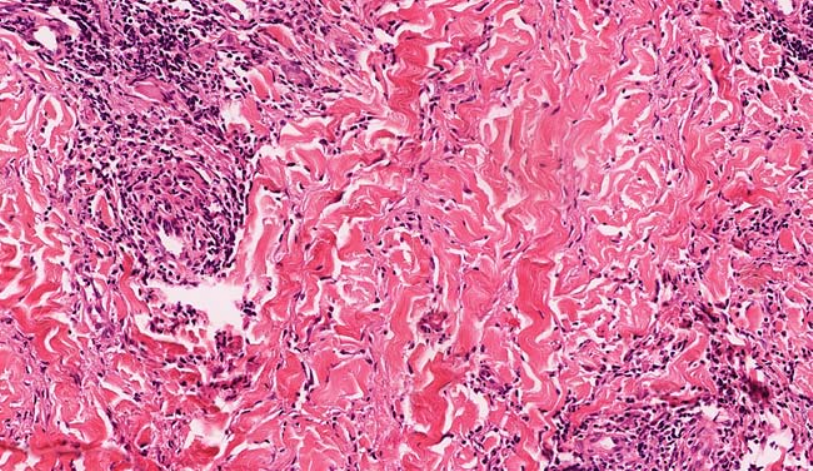}\\
\multirow{2}{*}{\includegraphics[width=0.20\textwidth]{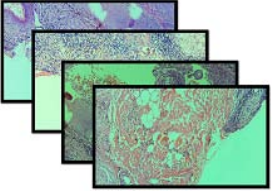}}&
\includegraphics[width=0.15\textwidth]{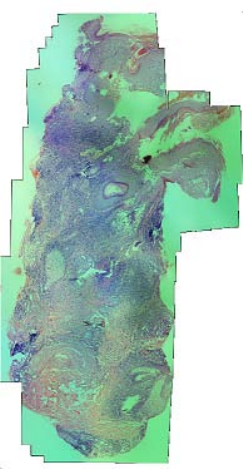}&
\includegraphics[width=0.15\textwidth]{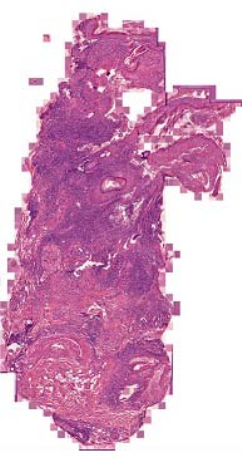}&
\includegraphics[width=0.15\textwidth]{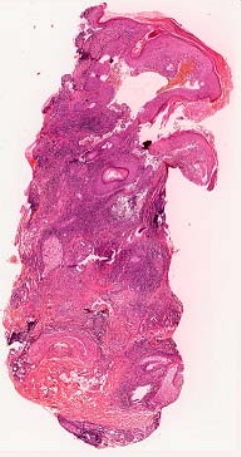}\\
&
\includegraphics[width=0.26\textwidth]{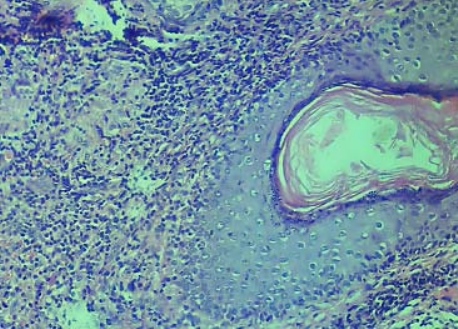}&
\includegraphics[width=0.26\textwidth]{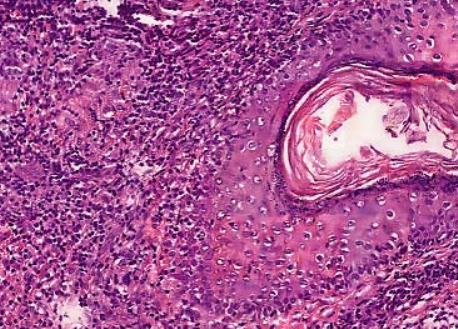}&
\includegraphics[width=0.26\textwidth]{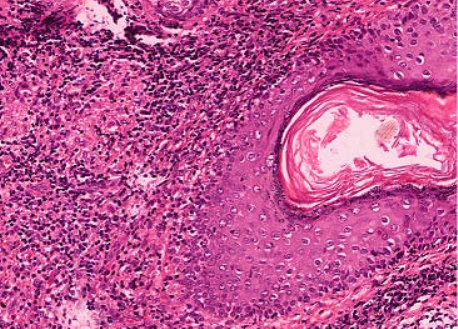}\\
\end{tabular}
\caption{Our pipeline for generating WSI-quality images from video and the corresponding scanned WSI for leishmaniasis cases.}
\label{fig:pipeline}
\end{center}
\end{figure*}

\section{Leishmaniasis and the curated dataset}
\textbf{Leishmaniasis background and histology presentation.} The World Health Organization (WHO) has selected cutaneous leishmaniasis as a neglected tropical disease (NTD) with growing, uncontrolled, and ignored infection affecting millions of people every year \cite{de2022cutaneous,world2020report}. It is unfortunately one of the world’s most neglected poverty‐related diseases, affecting the poorest people in developing countries and it is associated with risk factors like malnutrition, immune system deficiency, migration, inadequate education, illiteracy, gender inequality, and a shortage of services \cite{ahmad2022knowledge}. Furthermore, only eight countries in the world contribute to 90\% of leishmaniasis cases: Afghanistan, Algeria, Brazil, Iran, Pakistan, Peru, Saudi Arabia, and Syria \cite{sasidharan2021leishmaniasis}. It is a vector-borne infection caused by the protozoan parasite Leishmania. The vector is the female sand-fly. The lesions of cutaneous leishmaniasis vary in presentation ranging from a single self-limited skin lesion to multiple large destructive and ulcerated lesions on mostly the exposed parts of the body such as face, forearms, and lower legs \cite{bilgic2019cutaneous}. Skin biopsy is considered the gold standard method for confirmation of diagnosis. The presence of large epitheloid granulomas and small hematoxylinophilic LD bodies (that are round, uniform in appearance, intracytoplasmic and sometimes distributed around the outer rim of the vacuoles) is critical for diagnosis and confirmation of leishmaniasis \cite{handler2015cutaneous}. Large granulomas are visible at 10X but the smaller hematoxylinophilic bodies (around 3-4mm in size) are usually visible at 40X and can easily be missed if they are few in number, leading to misdiagnosis. Specialized dermatopathologists are the only ones who can diagnose these accurately based on patient history, clinical features, and laboratory diagnosis. It is thus critical to share the digitized slides of these cases from rural poor remote areas with dermatopathologists (mostly residing in urban centers) for quick diagnosis and accurate management of the patient.

\noindent
\textbf{Curated dataset.} Skin biopsies from 12 patients diagnosed with cutaneous leishmaniasis (spanning the disease differentiation spectrum) were acquired from Agha Khan University with Internal Review Board approval \# 2024-9038-28175. Six of these biopsies, stained with hematoxylin \& eosin (H\&E), were scanned using Huron's TissueScope digital scanner at 40X. A cheap-yet-clinical grade (\$200) microscope (BS-2020MD Digital Microscope from BestScope International Limited, China) with a built-in digital 1.3 megapixel CMOS USB camera and 1W 5-LED illumination with Halogen Lamp 6V/20W was then used to capture videos for all 12 skin biopsies at 10X resolution. To show generalizability of our whole-slide image creation workflow for other common pathologies beyond cutaneous leishmaniasis, we also acquired videos for core biopsies of breast, liver, duodenum, stomach, and lymph node.

\begin{table}[t!]
    \scriptsize
    \centering
    \caption{Model training parameters/configurations.}
    \label{tab:model_params}
    \setlength{\tabcolsep}{6pt}
    \begin{tabular}{|c|l|l|l|l|l|}
    \hline
    Model & Upsampling Layer & Padding Type & Generator Loss & LR & Padding\\
    \hline
    \#1 & transposed conv & zero & vanilla + L1 & 0.0002 & zero\\
    \#2 & resize conv & zero & vanilla + L1 & 0.0008 & zero\\
    \#3 & transposed conv & reflect & vanilla + L1 & 0.0008 & reflect\\
    \#4 & resize conv & reflect & vanilla + L1 & 0.0008 & reflect\\
    \textbf{\#5} & \textbf{transposed conv} & \textbf{reflect} & \textbf{vanilla + L1 + VGG} & \textbf{0.0004} & \textbf{reflect}\\
    \#6 & resize conv & reflect & vanilla + L1 + VGG & 0.0004 & reflect\\
    \#7 & resize conv & reflect & LSGAN + L1 + VGG & 0.0004 & reflect\\
    \hline
    \end{tabular}
\end{table}

\section{New Slide Digitization Workflow}

\subsection{Video creation guidelines and stitching}
Pathologists were instructed to record videos using a methodical and consistent movement over the slide, such as starting at one corner and sweeping complete rows, moving down a step and doing the sweep in the opposite direction, all the while pausing for a second between frames with at least 20--30\% overlap between adjacent frames. Following these instructions, the videos were recorded at 10X resolution for all the core biopsies, spanning 2--8 mins; we focus on core biopsies since these dominate the bulk of diagnosis work across diseases and these are fastest to sweep with video capture.

For stitching, we automatically extract the static (or ``pause'') frames using optical flow. A de-duplication step is run to discard the same frames. With the prior knowledge that these frames are contiguous with at least 20\% overlap, we use the open-source Python OpenStitching library (\url{https://github.com/OpenStitching/stitching}) to then stitch the extracted frames recursively in batches of 30--50 (stitched frames merged into super-stitched frames till a single WSI is generated). The recursive stitching brought down the time for WSI creation to a few minutes (in the order of 5--10 mins) from few hours with the naive stitching approach, which performs image registration between all frame pairs. 

\subsection{Image-to-image translation}
Once the images were stitched, we performed affine co-registration between the 6 corresponding video-stitched WSIs and the scanned WSIs. Two of these co-registered datasets were used to train a conditional generative adversarial network (pix2pix) with the remaining 4 used for testing; the 6 cases without scanned WSIs were used for additional validation and qualitative assessment by an expert pathologist. In total, 1408 image pairs were used for training and 453 for validation. For the generator, we used the standard 9-block ResNet module. We opted for the standard transposed convolution layers after trying checkerboard-artifact-mitigating alternatives such as resize-convolution, which in our use case introduced new artifacts such as gray spots and larger color variation. Reflect padding was used to reduce artifacts at the borders of the generated images and the VGG loss in the generator below dramatically improved the visual quality of the generated tiles. The discriminator was the 4-layer PatchGAN recommended in the original pix2pix paper \cite{isola2017image}. When feeding images into the discriminator, low-quality video patches were fed and high-quality patches were the generated condition to the ground truth high-quality scanned patch. 

\begin{table}[t]
    \scriptsize
    \centering
    \caption{Performance metrics on the test set.}
    \label{tab:metrics}
    \setlength{\tabcolsep}{6pt}
    \begin{tabular}{|c|c|c|}
    \hline
    Model & SSIM & PSNR\\
    \hline
    \#1 & 0.361 $\pm$ 0.033 & 11.078 $\pm$ 1.079\\
    \#2 & 0.363 $\pm$ 0.034 & 11.144 $\pm$ 1.546\\
    \#3 & 0.362 $\pm$ 0.039 & 11.090 $\pm$ 1.445\\
    \#4 & 0.376 $\pm$ 0.044 & 11.619 $\pm$ 1.604\\
    \textbf{\#5} & \textbf{0.317} $\pm$ \textbf{0.054} & \textbf{10.733} $\pm$ \textbf{1.221}\\
    \#6 & 0.319 $\pm$ 0.064 & 10.587 $\pm$ 1.431\\
    \#7 & 0.424 $\pm$ 0.041 & 11.167 $\pm$ 1.450\\
    \hline
    \end{tabular}
\end{table}

\begin{figure}[t]
\centering
\begin{tabular}{cccc}
\includegraphics[width=0.24\textwidth]{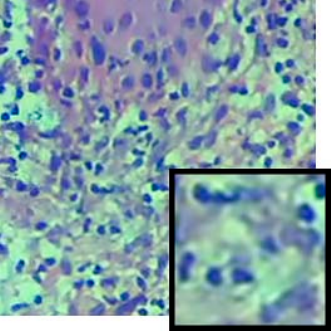}&
\includegraphics[width=0.24\textwidth]{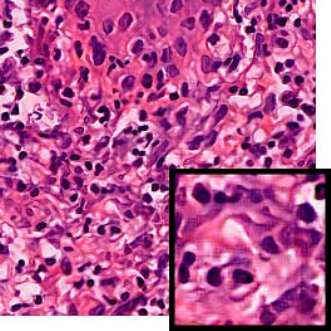}&
\includegraphics[width=0.24\textwidth]{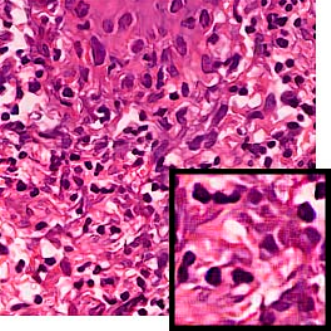}&
\includegraphics[width=0.24\textwidth]{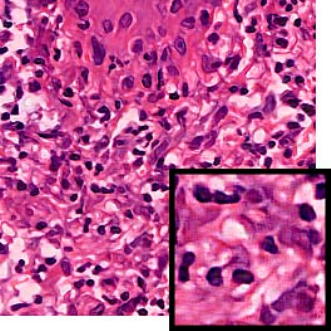}\\
Video & Model \#1 & Model \#2 & Model \#3\\[6pt]
\includegraphics[width=0.24\textwidth]{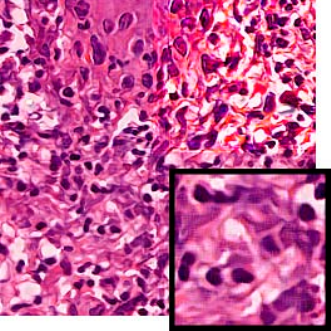}&
\includegraphics[width=0.24\textwidth]{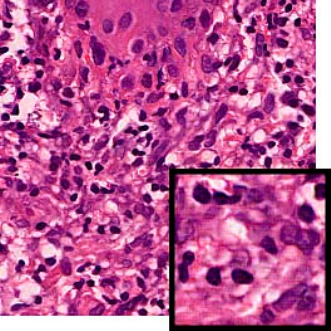}&
\includegraphics[width=0.24\textwidth]{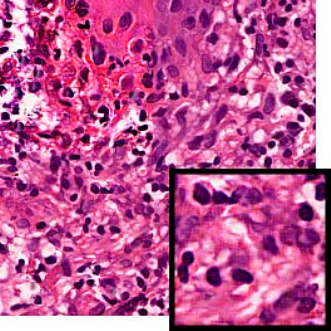}&
\includegraphics[width=0.24\textwidth]{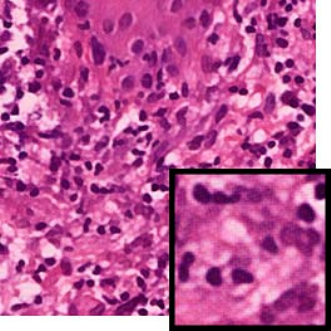}\\
Model \#4 & \textbf{Model \#5} & Model \#6 & Model \#7
\end{tabular}
\caption{Visual comparison of our seven trained models and the original video data.  Model \#5 gives the best visual results and is used for our final results.}
\label{fig:comp_models}
\end{figure}

In our application, the vanilla discriminator performed better than LSGAN discriminator for more realistic image generation. For the generator, a combination of mean-squared adversarial loss, VGG loss, and L1 loss was used with all given weights = 1. In line with Deblur-GAN-v2 \cite{kupyn2019deblurgan}, we also observed that the best generator was not the one with the best peak signal/noise ratio and structural similarity index measure scores; in fact a different model (Model \#5, using reflect padding + VGG, in Table \ref{tab:metrics}) generated the best visual image quality. Ablation study and details of different training configurations are given in Figure \ref{fig:comp_models} and Tables \ref{tab:model_params} and \ref{tab:metrics}. We implemented the model using PyTorch 1.10.2 framework. Adam optimizer was used to update gradients. The model was trained for 400 epochs with the learning rate set to 0.004 for the first 200 epochs and decays linearly. With batch size 16, the training took 84 hours on 2 A100 (40 GB) GPUs. The average runtime for training was 60 secs per epoch and for inference, it was 0.26 per tile. For memory footprint, training used 32.5 GB GPU memory while inference used 1.9 GB GPU memory.

\begin{figure*}[t]
\begin{center}
\setlength{\tabcolsep}{1pt}
\begin{tabular}{cccccc}
Input & DeblurGAN\cite{kupyn2019deblurgan} & SRGAN\cite{ledig2017photo} & ESRGAN\cite{wang2021real} & \textbf{Ours} & Scanned\\
\includegraphics[width=0.16\textwidth]{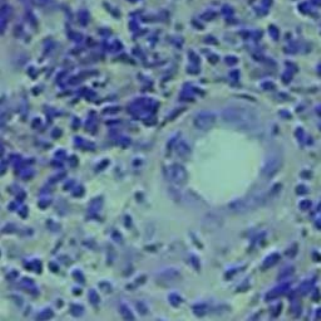}&
\includegraphics[width=0.16\textwidth]{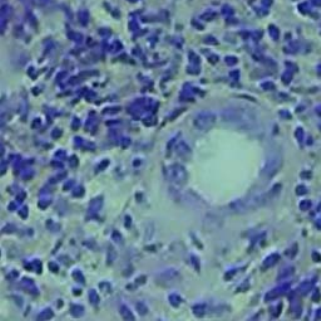}&
\includegraphics[width=0.16\textwidth]{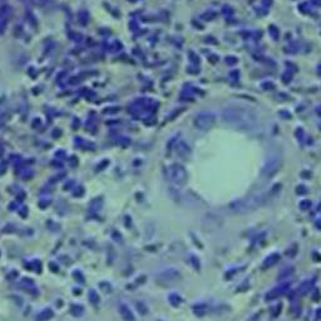}&
\includegraphics[width=0.16\textwidth]{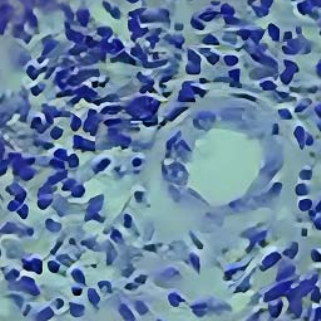}&
\includegraphics[width=0.16\textwidth]{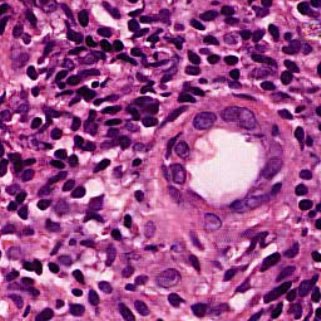}&
\includegraphics[width=0.16\textwidth]{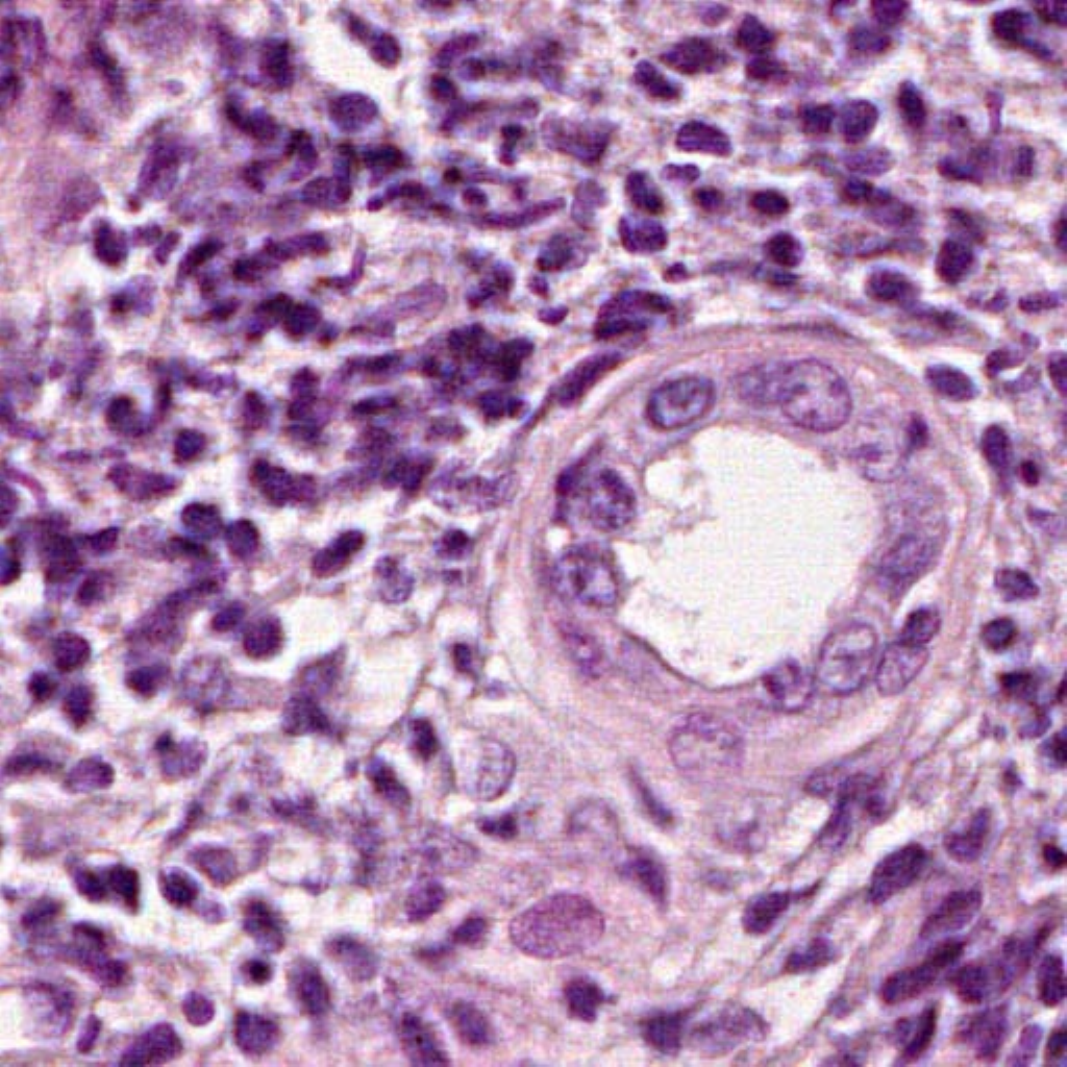}\\

\includegraphics[width=0.16\textwidth]{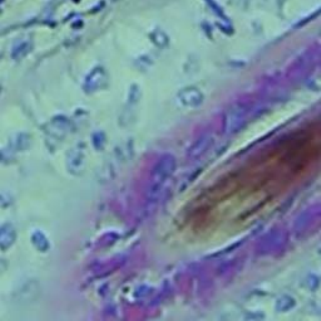}&
\includegraphics[width=0.16\textwidth]{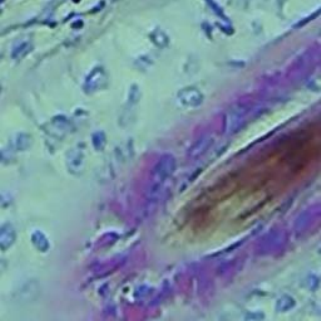}&
\includegraphics[width=0.16\textwidth]{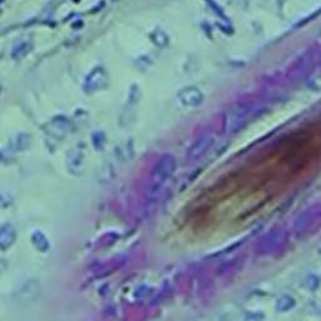}&
\includegraphics[width=0.16\textwidth]{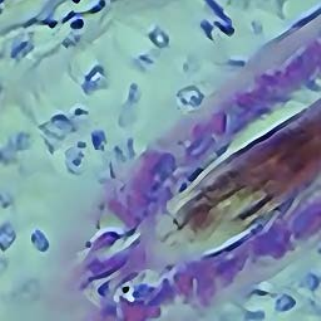}&
\includegraphics[width=0.16\textwidth]{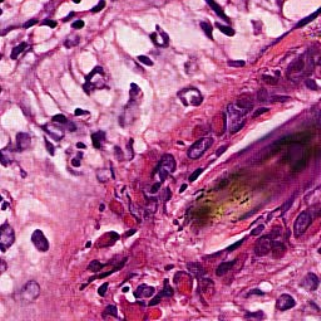}&
\includegraphics[width=0.16\textwidth]{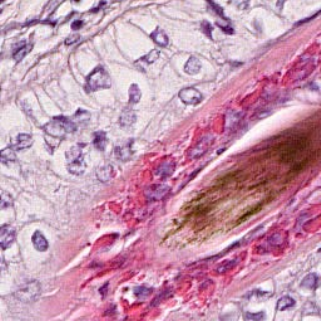}\\

\end{tabular}
\caption{Comparison with other image restoration methods.}
\label{fig:comp_others}
\end{center}
\end{figure*}

\section{Results and Discussion}
We have compared our approach against three image restoration approaches: DeblurGAN-v2 \cite{kupyn2019deblurgan}, SRGAN \cite{ledig2017photo}, and Real-ESRGAN \cite{wang2021real}. DeblurGAN corrects for Gaussian and motion blur (similar to Restore-GAN) where as SRGAN and Real-ESRGAN upsample images to 4X similar to what we are doing in this work (10X to 40X). As shown in the comparisons below in Figure \ref{fig:comp_others}, our model achieved better performance against these methods when run on real microscope video imagery.  In particular, DeblurGAN and SRGAN show little to no visual improvement to the input images.  Real-ESRGAN sharpens the image where there is sufficient contrast, but does not restore or enhance areas of low contrast.

The key difference between these other methods and our method is that we train on image data captured from real low quality videos and corresponding high quality scanned images. These videos contain natural variations in focus and illumination, which cause real world blur and brightness/contrast changes across different frames. Other models are trained on synthetic Gaussian and motion blur, which are not representative of the more complex distortions seen in real world data. Their training data also starts as high quality data, while we train with true lower quality image data that contains all defects found in real world video captured from microscopes.

\begin{figure}[t]
\centering
\begin{tabular}{ccc}
\includegraphics[height=0.189\textwidth]{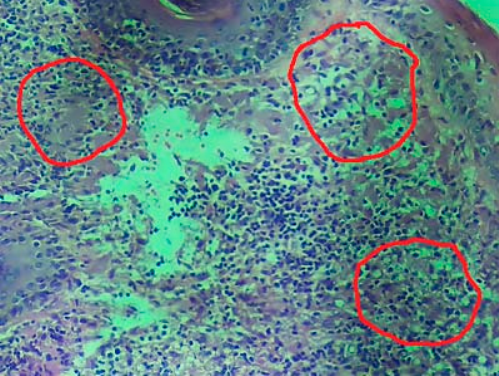}&
\includegraphics[height=0.189\textwidth]{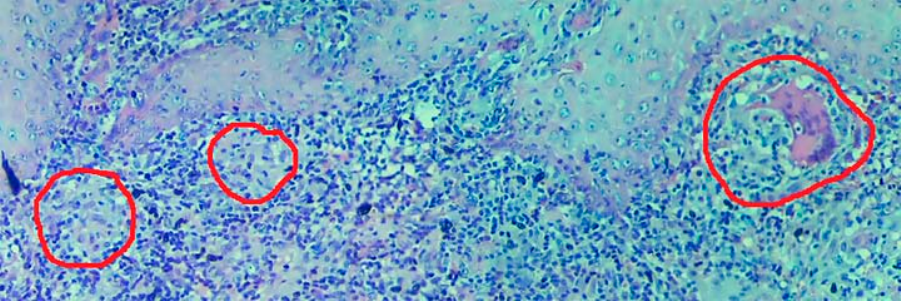}&
\includegraphics[height=0.189\textwidth]{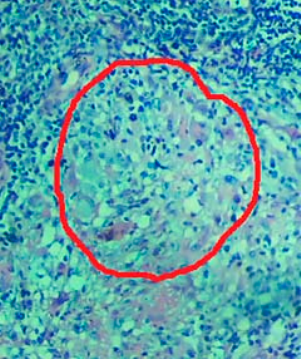}\\
\includegraphics[height=0.189\textwidth]{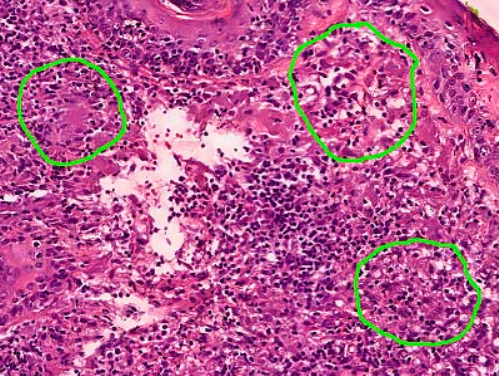}&
\includegraphics[height=0.189\textwidth]{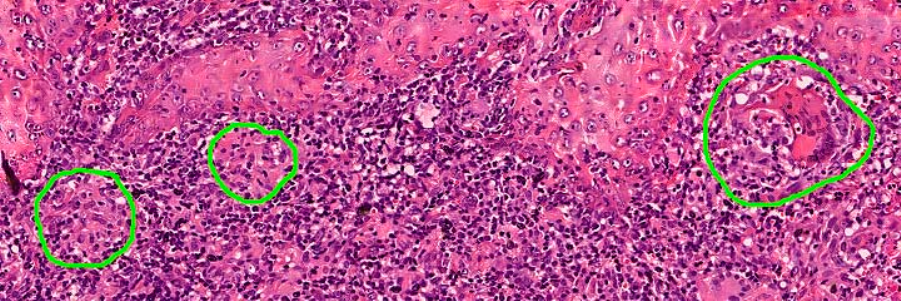}&
\includegraphics[height=0.189\textwidth]{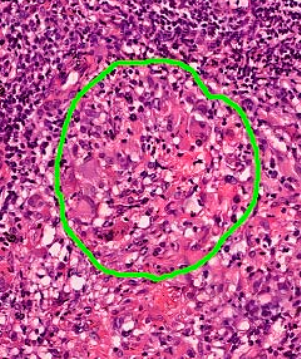}\\
\end{tabular}
\caption{Annotated granulomas identified in our stitched image from video (top row) and our generated WSI-quality (bottom row) images.}
\label{fig:annotate_granuloma}
\end{figure}

\begin{figure}[t]
\centering
\begin{tabular}{cc|cc|cc|cc|cc}
\includegraphics[width=0.07\textwidth]{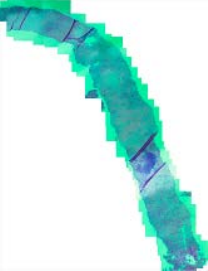}&
\includegraphics[width=0.07\textwidth]{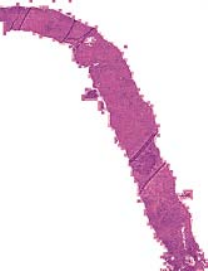}& 
\includegraphics[width=0.1\textwidth]{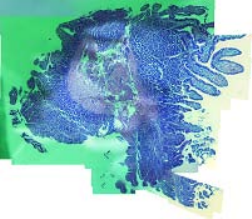}&
\includegraphics[width=0.1\textwidth]{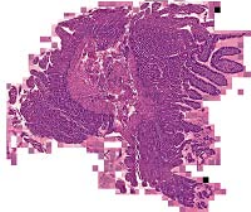}&
\includegraphics[width=0.12\textwidth]{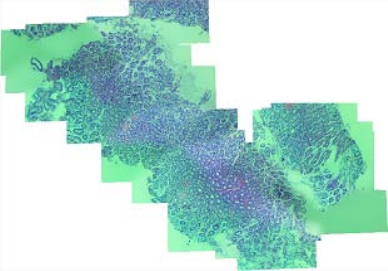}&
\includegraphics[width=0.12\textwidth]{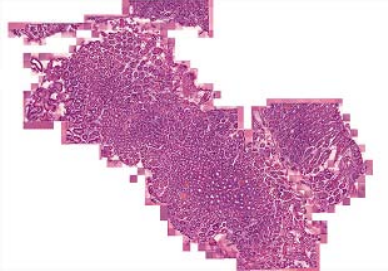}&
\includegraphics[width=0.05\textwidth]{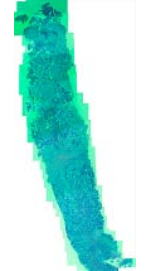}&
\includegraphics[width=0.05\textwidth]{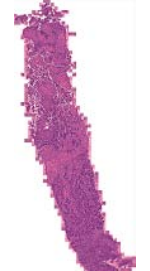}&
\includegraphics[width=0.08\textwidth]{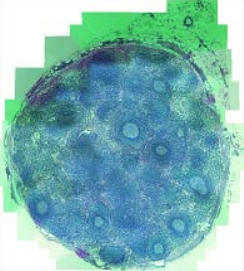}&
\includegraphics[width=0.08\textwidth]{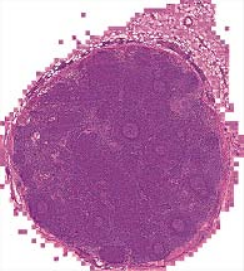}\\
\multicolumn{2}{c|}{Breast} & \multicolumn{2}{c|}{Duodenum} & \multicolumn{2}{c|}{stomach} & \multicolumn{2}{c|}{liver} & \multicolumn{2}{c}{lymph node}\\
\end{tabular}
\caption{Results on non-leishmaniasis core biopsies.}
\label{fig:results_other_sites}
\end{figure}

We demonstrate our WSI creation workflow on Leishmaniasis cases, as shown in Figures \ref{fig:pipeline} and \ref{fig:annotate_granuloma}. An expert pathologist reviewed the cases and annotated granulomas as well as LD bodies on the stitched images and our generated images in  Figure \ref{fig:annotate_granuloma}. To show generalizability of our approach, we also show results on core biopsies of breast, duodenum, stomach, liver, and lymph nodes (Figure \ref{fig:results_other_sites}).

\noindent
\textbf{Limitations.} Our current model results in significant artifacts at the boundary of the tissue. We will address these and the square artifacts at the borders more rigorously using bi-direction feature fusion GANs \cite{sun2023bi}. In this work, we used just one microscope which was the cheapest-yet-clinical grade microscope available on market (in developing regions). In the future, we will compare our stitching results on microscopes from different vendors with different resolution cameras. We will also explore using smartphone cameras to acquire videos rather than microscope cameras. We currently require pathologists to pause between frames (ideally with > 50\% overlap) to get the best-quality results. In the future, we will explore ways to handle stitching even from faster videos with minimal overlaps. We will also extend our stitching platform to immunohistochemistry-stained images where manual cell counting is tedious and more advanced deep learning approaches can play a critical role in more accurate, objective, reproducible, and faster protein expression quantification. Finally, we will develop deep learning algorithms for detecting granulomas and LD bodies specific to Leishmaniasis.

\noindent \textbf{Data use declaration and acknowledgment:} This study is not Human Subjects Research because it was a secondary analysis of results from biological specimens that were not collected for the purpose of the current study and for which the samples were fully anonymized. This work was supported by MSK Cancer Center Support Grant/Core Grant (P30 CA008748) and by James and Esther King Biomedical Research Grant (7JK02) and Moffitt Merit Society Award to C. H. Chung. It is also supported in part by the Moffitt’s Total Cancer Care Initiative, Collaborative Data Services, Biostatistics and Bioinformatics, and Tissue Core Facilities at the H. Lee Moffitt Cancer Center and Research Institute, an NCI-designated Comprehensive Cancer Center (P30-CA076292).


\begin{credits}
\subsubsection{\ackname} This research was funded in part through the NIH/NCI Cancer Center Support Grant P30 CA008748.

\subsubsection{\discintname}
The authors have no competing interests to declare that are relevant to the content of this article.
\end{credits}

%
%
%

\end{document}